# Magnetic-field controlled organic spintronic memristor for neural network computation

Tongxin Chen[1], Yinyu Nie[2], Yafei Hao[1,5], Shengchun Shen[3], Jiajun Pan[4], Xiaoguang Li[3], Yuan Lu[1*]

[1]*Université de Lorraine, CNRS, Institut Jean Lamour, F-54000 Nancy, France*
[2]*Technical University of Munich, Arcisstraße 21, 80333 München, Germany*
[3]*Hefei National Research Center for Physical Sciences at the Microscale, Department of Physics, University of Science and Technology of China, Hefei 230026, China*
[4]*University of Lorraine, Laboratoire Lorrain de Recherche en Informatique et ses Applications (LORIA), UMR 7503, 54506 Vandoeuvre-les-Nancy, France*
[5]*Department of Physics, Zhejiang Normal University, Jinhua 321004, China*

Corresponding authors*: *yuan.lu@univ-lorraine.fr*

**Abstract**

Memristors are emerging as key electronic components that retain resistance states without power. Their non-volatile nature and ability to mimic synaptic behavior make them ideal for next-generation memory technologies and neuromorphic computing systems inspired by the human brain. In this study, we present a novel organic spintronic memristor based on a $La_{0.67}Sr_{0.33}MnO_3$ (LSMO)/poly(vinylidene fluoride) (PVDF)/Co heterostructure, exhibiting biologically inspired synaptic behavior. Driven by fluorine atom migration within the PVDF layer, the device demonstrates both long-term depression (LTD) and long-term potentiation (LTP) under controlled electrical polarization. Distinctively, the resistance states can also be modulated by an external magnetic field via the tunneling magnetoresistance (TMR) effect, introducing a non-electrical means of tuning synaptic plasticity. This magnetic control mechanism enables multi-state modulation without compromising device performance or endurance. Furthermore, convolutional neural network (CNN) simulations incorporating this magnetic tuning capability reveal enhanced pattern recognition accuracy and improved training stability, especially at high learning rates. These findings underscore the potential of organic spintronic memristors as high-performance, low-power neuromorphic elements, particularly suited for applications in flexible and wearable electronics.

**Introduction**

Memristors have captured considerable attention as promising components in advanced computing due to their unique ability to remember past resistance states,[1] making them ideal for applications in non-volatile memory,[2] neuromorphic computing,[3] and threshold logic.[4] Their memristive behavior typically relies on physical mechanisms that enable a reversible transition between high and low resistance states, such as formation of metallic filament,[5,6] agglomeration,[7] migration of oxygen vacancies,[8,9] charging or discharging of nanoparticles in the conduction channel,[10] metal substitution in 2D materials,[11,12] ferroelectricity of tunnel barriers,[13,14] magnetic domain motion in magnetic tunnel junctions (MTJs),[15] and mechano-gated iontronic piezomemristor.[16]

For neuromorphic computing, memristors can serve as artificial synapses that enable energy-efficient and brain-inspired learning. Unlike traditional von Neumann architectures, memristor-based systems can process and store data in the same location, reducing latency and energy consumption in artificial intelligence (AI) applications. Their ability to exhibit synaptic plasticity makes them ideal for implementing spiking neural networks (SNNs), which mimic biological neurons for real-time learning and pattern recognition. However, one of the major challenges in deploying memristors for neuromorphic systems is that their endurance (frequent switching) can cause degradation in resistive states, leading to performance instability over time. Additionally, while memristors are inherently low-power devices, their power consumption can rise significantly in large-scale neuromorphic architectures due to cumulative switching events across massive networks. Overcoming these challenges requires advancements in materials engineering and device optimization to enhance the memristor's performance, durability, and energy efficiency for long-term neuromorphic applications.

Tunneling magnetoresistance (TMR) arises in magnetic tunnel junctions where two ferromagnetic (FM) electrodes are separated by an ultrathin insulator. Because electron tunneling the insulating barrier is spin dependent, the junction resistance changes with the relative magnetization orientation of the electrodes: it is typically lower for the parallel state and higher for the antiparallel state. Combining the TMR effect with electrically driven memristors paves the way for the development of novel spintronic memristors. In these devices, resistance changes are influenced not only by electric-field-driven memristive mechanism but also on the magnetization configurations of the FM electrodes. TMR introduces an additional degree of freedom for controlling the plasticity characteristics of memristor-based artificial synapses, which is essential for emulating the morphological alterations observed in biological synapses.

Although such spintronic memristors have been demonstrated using multiferroic tunnel junctions (MFTJs)—where a ferroelectric barrier is sandwiched between two ferromagnetic electrodes—their potential advantages for neuromorphic computing remain largely unexplored compared to conventional electrically driven memristor-based neural networks. Recent studies have begun to reveal their promise: Huang *et al.* reported an MFTJ based on $La_{0.7}Sr_{0.3}MnO_3/BaTiO_3/La_{0.7}Sr_{0.3}MnO_3$, where magnetoelectric coupling enables continuously tunable spin polarization and synaptic plasticity forms can be manipulated.[17] Yang *et al.* demonstrated $BaTiO_3/CoFe_2O_4$-based FTJs exhibiting high ON/OFF ratios, reversible tunneling magnetoresistance, and successful pattern recognition with over 97% accuracy in a crossbar neural network.[18] These works highlight the potential of MFTJs for low-power, non-volatile, and functionally rich artificial synapses. However, a systematic evaluation of their full potential for neural network performance remains to be conducted.

In this work, we fabricated $La_{0.67}Sr_{0.33}MnO_3$(LSMO)/poly(vinylidene fluoride)(PVDF)/Co organic spintronic memristor and investigated their memristive behavior through voltage-induced polarization and magnetic-field-dependent TMR effects. By controlling the variables, we prove that it is capable to control the junction resistance by voltage pulse and magnetic field simultaneously.

The additional tunability by magnetic field provides extra resistance states in individual memristor devices, leading to an enhanced performance in neuromorphic computing with higher recognition accuracy and faster training process, which is suitable for high performance and energy-efficient AI applications. Moreover, the prospect of fabricating large-scale organic layers through low-cost, solution-based Langmuir-Blodgett processes makes these developments highly attractive for wearable, flexible, and implantable device applications.[19]

**Results and discussion**

*Magneto-transport characterizations of LSMO/PVDF/Co junctions*

**Figure 1a** is the schematic illustration of the LSMO/PVDF/Co junction. The ultrathin PVDF (3 layers) was deposited on the prepatterned LSMO substrate using Langmuir Blodget (LB) method.[20,21] In this work, we fabricated small junctions ($10\times10\mu m^2$) with UV lithography technique which has been described in our previous work.[21] Although PVDF and its copolymers are well-known for their superior ferroelectric properties[22] due to the formation of dipoles between the positively charged hydrogen ions ($H^+$) and the negatively charged fluorine ions ($F^-$),[23] the distinctive memristive mechanism in our organic memristor is the voltage driven motion of fluorine (F) atoms in the junction,[21] as illustrated in **Figure 1b**. After annealing at 120°C, the fluorine components in the thin PVDF layer are completely dissociated by the LSMO bottom electrode (see details in Ref.[21] and **Methods**). Upon electrical polarization, the fluorine components can be driven either to the LSMO/PVDF interface or into the top CoO/Co layer. When the junction is positively polarized (upper panel of **Fig. 1b**), the voltage-driven F atoms enter the LSMO side and the effective tunneling barrier thickness is considered as the sum of PVDF and CoO layer, resulting in a high resistance state. While the junction is negatively polarized (bottom panel of **Fig. 1b**), the F atoms enter the CoO layer. The F-doped CoO becomes conductive and therefore only PVDF layer is considered as the tunneling barrier, leading to a low resistance state. Please find more morphology, interfacial structure and chemical characterizations in **Notes 1 and 2** (Supporting Information (SI)).

**Figure 1c** shows the *I-V* measurement obtained when the junction is unpolarized, polarized by +2 V (1 s duration) pulse and polarized by -2 V (1 s) pulse, respectively. The different *I-V* curves prove that the incorporation of F into different interfaces can tune the junction resistance by changing the effective tunneling barrier thickness. In **Figure 1d**, we demonstrate the resistance evolution of our junction under successive change of polarization voltages. We observe a gradual increase (decrease) of resistance with increasing positive (negative) polarization voltage. There are two resistance plateaus which are defined as high resistance state (HRS, $R_\text{H}$) and low resistance state (LRS, $R_\text{L}$), with a huge resistance change ratio (RCR, $(R_\text{H}-R_\text{L})/R_\text{L}\times 100\%$) up to $1.1\times 10^4\%$. Between these two resistance states, it is possible to obtain various resistance states by changing the pulse width and amplitude of the polarization voltage. The changeable resistance states form the basis of memristor behavior.

The unique advantage of our memristor is the significant TMR effect,[21] which is distinct from other reported PVDF-based memristors driven by ferroelectricity.[24,25] The TMR effect is a quantum mechanical phenomenon observed in MTJs,[26–28] where the electrical resistance of the junction depends on the relative alignment of magnetizations in two ferromagnetic layers separated by an ultrathin insulating barrier. As shown in **Figure 1e**, our junction exhibits a strong negative TMR following the negative polarization $V_\text{p}$=-2.4 V (1 s). Under a large magnetic field (2 kG), where the magnetizations of Co and LSMO are aligned parallel (P), the junction is in a high resistance state. Due to the smaller coercivity of LSMO compared to Co, the LSMO layer switches its magnetization first at a small opposite field (90 G), resulting in an antiparallel (AP) configuration where a lower resistance state is observed. The TMR ratio, defined as $(R_{AP}-R_P)/R_{AP}\times 100\%$, where $R_P$ and $R_{AP}$ are the junction resistances in the P and AP states respectively, is calculated to be -43.9%.

In our junction, the motion of F atoms plays an important role to tune the TMR effect. When organic molecules are in contact with ferromagnetic metals, the organic molecule orbits would hybridize with the spin-split bands of ferromagnetic metal, leading to a so called "spininterface" with highly efficient spin filtering effect.[29,30] As the F atoms are electrically driven to either LSMO or CoO interface, the hybridization at interface would be remarkably altered, resulting in tunable sign and amplitude of TMR effect.[30–33] In **Figure 1f**, we demonstrate the change of TMR sign under different electrical polarizations. The junction exhibits a positive TMR at -0.8 V polarization and a negative TMR at -1.3 V polarization. The TMR sign switching occurs at around -1.2 V polarization. The sign change in TMR is attributed to a reversal in spin polarization at the PVDF/CoO interface, depending on the presence or absence of F incorporation.[21] Notably, the TMR ratio is significantly reduced in these intermediate resistance states compared to the fully polarized state shown in **Fig. 1e**., due to the coexistence of F at both the CoO/PVDF and LSMO/PVDF interfaces (see **Note 2 in SI**). This results in a rapid decrease in TMR, primarily caused by a reduction in spin polarization at the LSMO/PVDF interface due to the F insertion.[21] The tunability of TMR through electrical polarization provides the flexibility to modulate neural synaptic functions via control of the TMR effect.

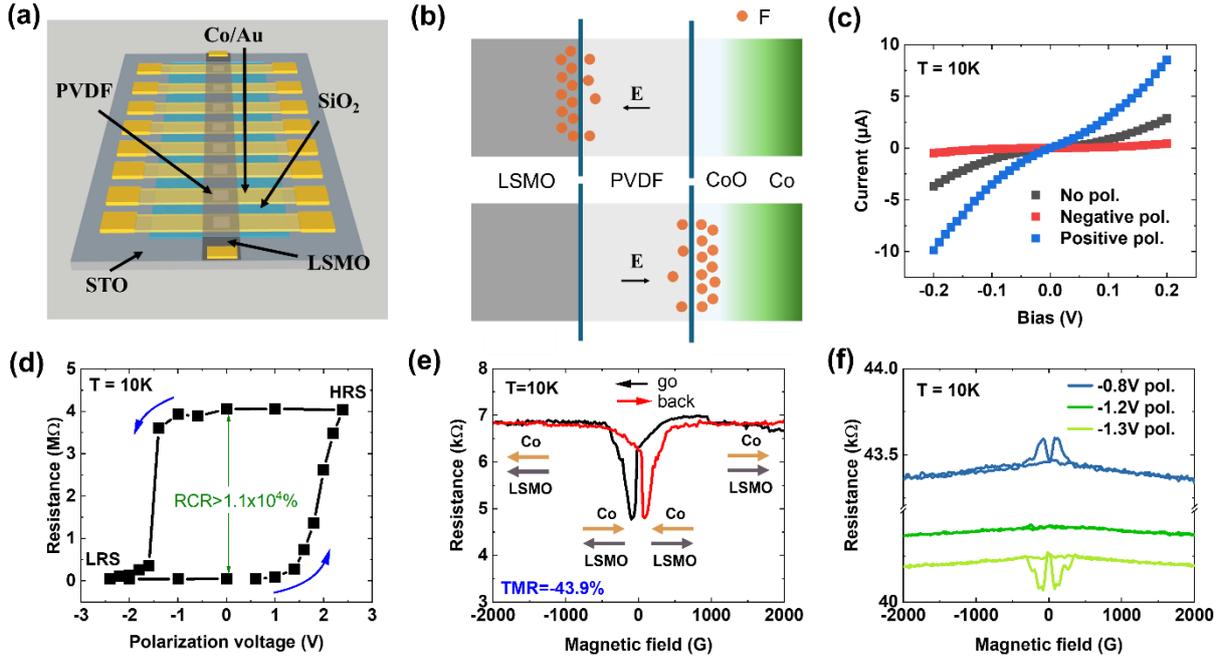

**Figure 1:** (a) Schematics of LSMO/PVDF/Co/Au memristor. The junction size is 10×10μm$^2$. (b) Schematics of F atom motion under different electric polarization fields. The electric polarization can either drive the F atoms to the bottom LSMO side (positive polarization) or to the top CoO layer (negative polarization). (c) *I-V* measurement of non-polarized, positive and negative polarized states at 10K. (d) Successive evolution of the junction resistance under a series of electric polarization with pulse duration of 1 s at *T* = 10 K. Resistance was measured under $V_{mes}$ =+10 mV. (e) TMR measured under $V_{mes}$ = +10 mV at *T* = 10 K after negative polarization with $V_P$ = -2.4 V (1 s). Black and red arrows indicate the magnetic field sweeping direction. Inserts: four resistance states associated with different magnetization orientations of Co and LSMO electrodes. (f) Change of TMR sign and amplitude with successive -0.8 V, -1.2 V and -1.3 V electrical polarizations at *T* = 10 K with pulse duration of 1 s. The TMR curves are measured under $V_{mes}$ =+10 mV.

*Magnetic field tunning of synaptic function*

In the previous section we present the tunable resistance states with 1 s polarization pulse. However, the speed to modify the resistance states is critical in real applications. Therefore, in **Figure 2a** we demonstrate the ability to continuously adjust the junction resistance by a series of negative (-1.5 V, 10 ms) and positive (0.8 V, 10 ms) pulses, applied when the junction is at a relatively low resistance state. As the width of polarization pulse is much shorter, the junction resistance could not be fully switched to LRS or HRS, resulting in a gradual change in resistance, different from what we demonstrate in **Fig. 1d**. The positive and negative pulse amplitude were carefully chosen to obtain stable and reversible resistance switching, with more details in **Figure S3** (**Note 3 in SI**). Negative pulses (-0.8 V) gradually decrease the junction resistance (i.e., increase the conductance), while positive pulses (+1.5 V) increase the resistance (i.e., decrease the conductance). These results closely resemble the behaviors of long-term potentiation (LTP) and long-term depression (LTD), respectively, mimicking the processes of neuronal potentiation and depression.[34] This plasticity-tunable artificial synapse makes it suitable for applications such as neuromorphic computing.[35]

Different from conventional neuromorphic devices, we can tune the memristor resistance not only by applying voltage pulses but also by magnetic field due to the TMR effect. We have obtained positive TMR in both -0.8 V and +1.5 V polarization conditions. When the magnetization of the two FM electrodes (Co and LSMO) are parallel (antiparallel), the memristor has a low (high) resistance state. In **Figures 2b** and **2c**, we demonstrate the evolution of the TMR signal as a function of pulse number during the LTP and LTD process, respectively. In both circumstances, the sign of TMR effect conserves to be positive while increasing the pulse number. This suggests that it is possible to manipulate memristor resistance independently by two parameters either electrical pulse or magnetic field.

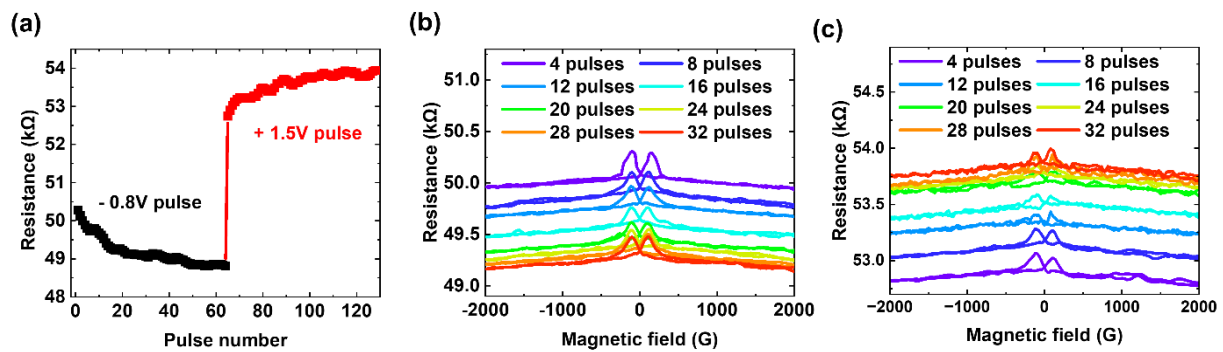

**Figure 2:** (a) LTP curve with -0.8 V pulse polarization (black) and LTD curve with +1.5 V pulse polarization (red). The measurement was carried at LRS with pulse duration of 10 ms. (b) Evolution of TMR curves as the number of -0.8 V pulse increases. (c) Evolution of TMR curves as the number of +1.5V pulse increases. For both measurements in (b) and (c), the device was first initialized with 64 pulses of +1.5 V/-0.8 V polarization and then polarized with reversed pulses.

To further explore the tuning possibility by magnetic field, we conducted another LTD/LTP measurement with +2 V (10 ms) and -1.3 V (10 ms) pulses. The amplitude of the pulse pair was carefully chosen to maximize the resistance variation, while **Figure S4** (**Note 3 in SI**) shows examples of improperly functioning amplitude pairs. Different from previous measurement, the increased negative pulse amplitude can switch the sign of TMR to be negative (**Figure 3a**). In **Figure 3b** we show the evolution of the junction's resistance under repeated cycles of positive and negative polarizations. Upon confirmation of reversible LTD and LTP behavior of our device, we measured the LTD and LTP curves under different magnetic fields ranging from 0 to 350 G (**Figures 3c** and **3d**, respectively). Under all conditions, our devices exhibit good LTD and LTP behaviors. More importantly, when comparing the resistance at the same pulse number, its evolution closely follows the shape of the TMR curve. This confirms our conclusion that the memristor resistance can be independently controlled by electrical pulses and magnetic field, in both the positive and negative TMR region.

Our device exhibits non-volatile plasticity, with no observable short-term potentiation or depression (STP/STD). Once set by an electrical stimulus, the resistance remains stable and does not relax without an opposite-polarity pulse, due to voltage-driven fluorine migration at the LSMO/PVDF and PVDF/CoO interfaces (**Fig. 1b**), which stabilizes the device in either a HRS or LRS. Fluorine motion becomes immobilized post-switching, ensuring long-term retention (**Note 4 in SI**). Although we did not explicitly measure resistance stability under a static magnetic field, stability is confirmed by the overlapping resistance values at $H$=+2 kG before and after a full ±2 kG magnetic field sweep (~20 min per cycle), as shown in **Fig. 2b,c**. This indicates no resistance drift under constant field with time.

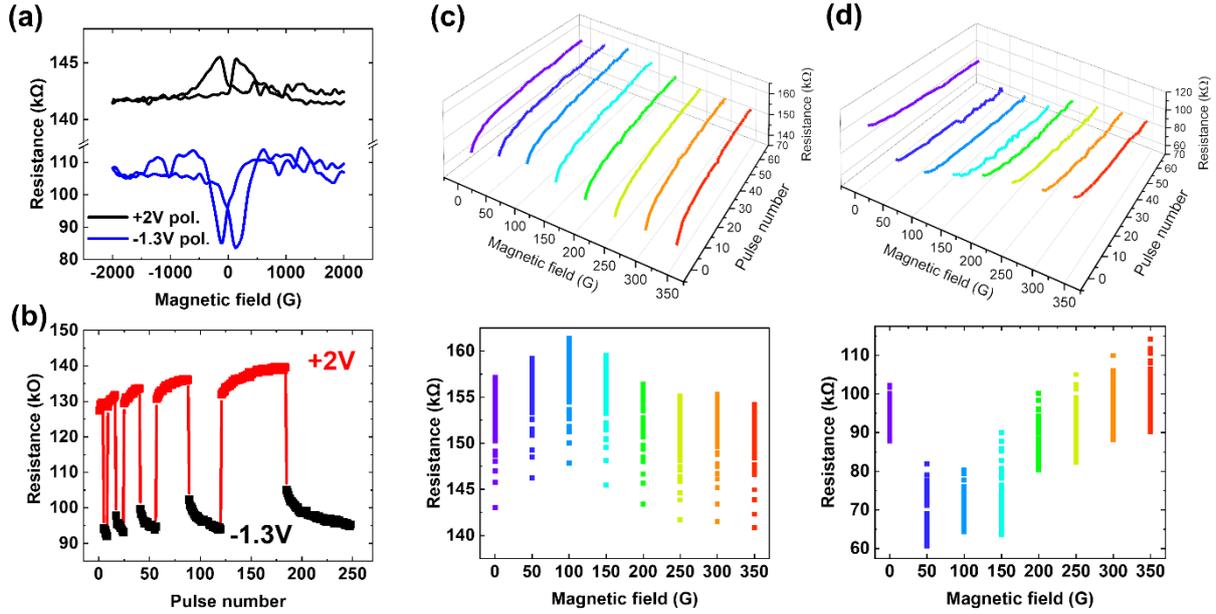

**Figure 3:** (a) TMR curves measured under $V_{mes}$ =+10 mV at 10 K after +2 V and -1.3 V polarizations (10 ms pulse). (b) Repeated LTD and LTP loops with successive +2 V pulse and -1.3 V pulse. Pulse duration is 10 ms. (c-d) Evolution of (c) LTD and (d) LTP curves as a function of magnetic field. All the curves were measured after electrical initialization with reversed polarization at -1.3 V or +2 V and magnetic initialization with -2 kG magnetic field. Under all magnetic fields, the junction resistance follows the LTD/LTP pattern as the pulse number increases. The lower panels of (c) and (d) show the evolution of LTD/LTP as a function of magnetic field. At the same pulse number, the junction resistance well follows the TMR curve trend.

## *Manipulation of synaptic plasticity with magnetic field*

Emulating spike-timing-dependent plasticity (STDP) in a memristor within a competitive Hebbian learning framework represents a promising step toward bridging the gap between biological and artificial neural systems.[36] The STDP is a biological learning rule that describes how the connection between neurons, also known as synaptic strength, is modified based on the relative timing of spikes between pre- and post-synaptic neurons. If the pre-synaptic neuron fires slightly before the post-synaptic neuron (Δt > 0), the synapse is strengthened. Conversely, if the pre-synaptic neuron fires slightly after the post-synaptic neuron (Δt < 0), the synapse is weakened. In both cases, the closer the timing of the spikes (Δt), the stronger the effect on synaptic modification, whether strengthening or weakening.[37]

In **Figure 4**, we demonstrate the STDP measurement with our device. Two voltage waveforms,

shown in **Figure 4a**, present pre- and post-synaptic spikes. Each spike consists of a pulse followed by an opposite smooth slope, and the two waveforms are of opposite polarity. The amplitude and width of the spikes are carefully designed to prevent a single spike from altering resistance by staying below a threshold ($V_{th}$) (Please see **Note 5 in SI** for details). When pre- and post-synaptic spikes overlap within a specific time (Δt), the waveform temporarily exceeds the threshold, modifying the synaptic connectivity.[38] **Figure 4b** illustrates the STDP measurement when the magnetizations of Co and LSMO are in P and AP configurations. Under both conditions, positive Δt strengthens (potentiates) the connection (i.e. conductance change Δ$G$ > 0), while negative Δt weakens (depresses) the connection (i.e. conductance change Δ$G$ < 0), closely replicating biological synaptic behaviors.[3] It is interesting to find that the conductance change ratio (Δ$G$/$G$) is larger in the AP state than the P state, which is contrary to the result reported by Huang *et al.*.[17] This could be due to the negative TMR effect in our device. The STDP results demonstrate that the neural synaptic response can be manipulated by altering the magnetization alignment of the FM electrodes through changing the magnetic field.

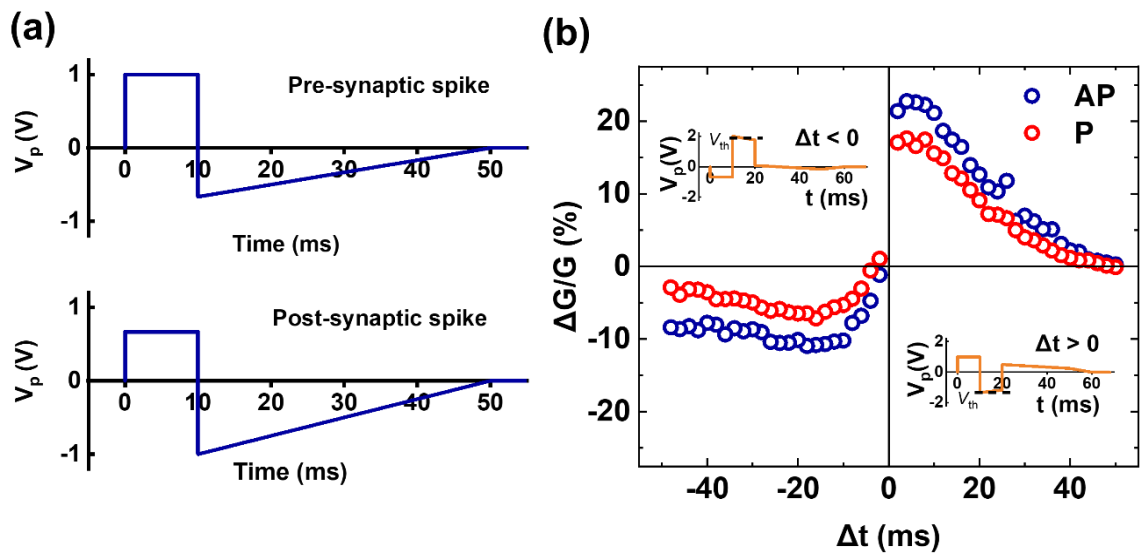

**Figure 4:** Measurement of STDP in the LSMO/PVDF/Co memristor. (a) pre- (upper) and post-synaptic spikes (lower) with a total length of 50 ms. The amplitudes of the rectangular voltage pulse and opposite slope are +1.0 and −1.5 V for the spike from pre-synaptic neuron, and +1.5 and −1.0 V for the spike from post-synaptic neuron, respectively. (b) Ratio of device conductance change (Δ$G$/$G$) as a function of time interval (Δt) between pre- and post-synaptic spikes. Insets:

Waveforms resulting from superposition of pre- and post-synaptic spikes when Δt<0 or Δt>0.

*Neural network architecture for image classification*

To evaluate the performance of our device in real neuromorphic computation tasks, we design one image classification network as displayed in **Figure 5a**. The entire architecture follows a classic convolutional pipeline, with two blocks of two dimensional (2-D) convolutional layers to process the input images into 2-D convolutional features (top row in **Figure 5a**), followed by two blocks of multi-linear perceptron (MLP) layers to predict the category score of each label for image classification.[39] In contrast to traditional convolutional neural networks which often use Rectified Linear Units (ReLU)[40] or alternative computation units (e.g., ELU, LeakyReLU, sigmoid, tanh) as the activation function,[41] our key design is using an activation layer based on our experimental results to activate the features from upstream layers to capture the non-linear property for image classification.

We demonstrate the architecture of our electrical-magnetic (EM) activation layer in **Figure 5b**. Similar to ReLU, our activation layer processes each entry of the input signals independently such that the output signals keep the same dimension as the input. For each element *x* in the input signals, we feed it with LTD and LTP curves (which are fitted via experiment results) and then generate the corresponding output pulse vector that represents the response of our device. By changing the input LTD and LTP curves, we can simulate the device performance under different conditions. Afterwards, we process the vector with an MLP layer to reduce its dimension to 1-D and place it back to the original location, so that we can accomplish an activation pass to capture the nonlinear feature from the input signals and keep the input dimension unchanged.

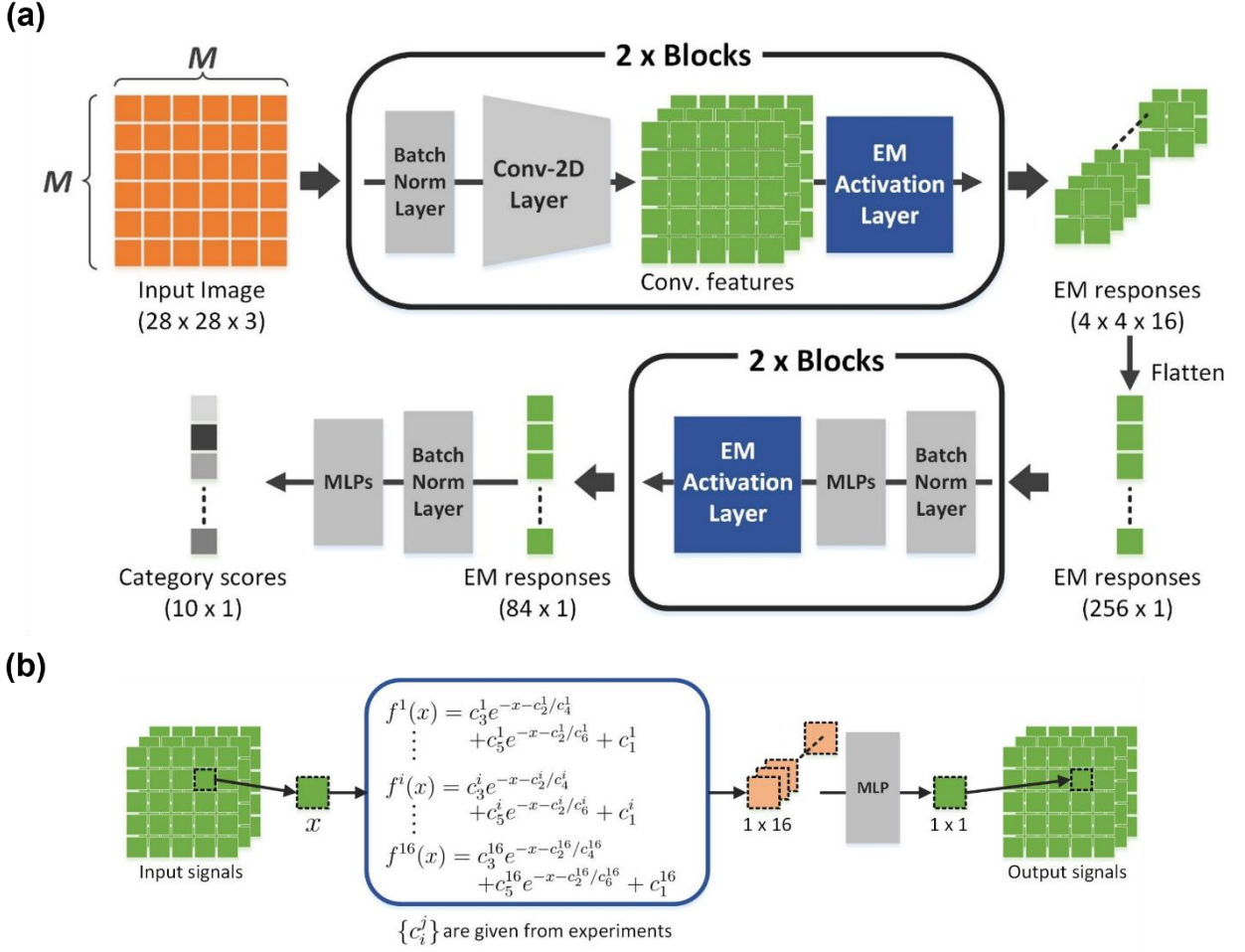

**Figure 5**: (a) Neuro network architecture for image classification. The network begins with a 28 × 28 × 3 input image processed by two convolutional blocks, each comprising a BatchNorm layer, a 2-D convolutional layer, and an electrical-magnetic (EM) activation layer to produce 4 × 4 × 16 EM responses. These feature maps are flattened into a 256 × 1 vector and passed through two fully-connected blocks (EM Activation → MLP → BatchNorm), producing 84 × 1 EM responses. A final MLP then generates 10 category scores. (b) Neural network design of EM activation layer. Inside each EM activation layer, each scalar input $x$ is transformed into a 1 × 16 vector via 16 distinct exponential-mixture functions. A small MLP then reduces this 1 × 16 vector back to a single output signal at each spatial location.

We designed three experiments to simulate different conditions. We first defined the baseline using the original version of CNN with ReLU activation function, which is widely used due to its simplicity and effectiveness.[41] The baseline represents the optimal neural network we can obtain by solely computer simulation. The second experiment (SET1) is to simulate a conventional artificial synapse device driven only by electrical pulses. We conduct the simulation by inputting our network with one set of LTD and LTP curves measured with zero magnetic field as the activation layer. The third experiment (SET2) aims to simulate the conditions under which more resistance states can be achieved in our device by tuning the magnetic field. In this case, we input 8 sets of LTD and LTP curves illustrated in **Figs. 3c** and **3d** as the EM activation layer into the neural network. In **Figure 6a** we compared the simulation results obtained from Baseline, SET1 and SET2. All the simulation results were obtained using the same condition (2E-5 learning rate). We obtained a baseline with about 90% recognition accuracy after 100 epochs, which is comparable to similar CNN networks reported in other works.[42] This proves that the network is suitable for such recognition tasks.

In SET1, we demonstrate the feasibility of using a memristor to implement the convolution layer. The input from image is converted into the pulse number applied on the memristor and 3 × 3 kernel convolution is done by addition of the resistance of 9 individual memristors. The integration of memristors into neural networks presents a transformative advantage by enabling highly efficient and biologically inspired computing. Different from traditional digital systems, memristors naturally store weights as analog resistance values, supporting continuous and precise update. This architecture is ideal for neural networks, which strongly rely on the weight storage and repeated matrix-vector multiplications, by allowing in-memory processing that significantly reduces the data movement and time delay between memory process and computation. Also, highly integrated memristor arrays enable highly efficient parallel computation architecture.

However, the accuracy of recognition in SET1 can only reach around 84% after 100 epochs, around 5% lower than the baseline. This is due to the convolution calculation with memristor uses a customized activation function (pulse *vs.* resistance curve) compared with ReLU activation in the baseline. In neural networks, the choice of activation function plays a critical role in determining the efficiency and success of learning. The ReLU activation function introduces non-linearity while allowing for fast computation and efficient gradient propagation. It avoids the vanishing gradient problem commonly associated with sigmoid or tanh activations, making it well-suited for deep networks. On the other hand, the customized activation function in memristor architecture involves a more complex formula, featuring exponential terms and multiple trainable coefficients. Although it precisely describes the physics phenomenon in the memristor, it also introduces several challenges. The use of exponential term makes the gradients become extremely small, leading to slow or stalled learning, especially in deeper layers. Moreover, the additional parameters may lead to overfitting, especially on small datasets.

Remarkably, our spintronic memristor with additional magnetic-field introduced resistance states (SET2) achieves approximately 89% accuracy after 100 epochs, closely approaching the baseline performance. The CNNs learn hierarchical feature and the non-linearity of the activation function play a crucial role in determining how efficiently the network can extract complex patterns. If one activation function is too linear, it may limit the representational power of the network. Thanks to the extra resistance states induced by the magnetic field in SET2, we have more data input into the activation layer than that in SET1. The extra data input in the activation layer improves the non-linearity, leading to the better recognition performance.

Additionally, the advantage of our memristor also appears when increasing the learning rate in the neural network. The learning rate is a hyperparameter that controls how much the model's weights are updated during the training process and determines the step size taken in the direction of the gradient when optimizing the loss function. If the learning rate is too slow, the model may converge too slowly or get stuck in a local minimum, while a too high learning rate may overshoot the optimal weights, leading to instability or divergence. Within a reasonable range, a faster learning rate is usually favorable as the faster convergence enables the model to reach an optimal solution quicker and reduce the training time. **Figure 6b** shows the comparison between a slow learning rate (2E-5) and a fast learning rate (1E-4) with SET2 data. With the higher learning rate, the neural network reaches 89% accuracy within 10 epochs, while with lower learning rate it requires 60 epochs to reach a similar accuracy. **Figure 6c** illustrates the average time of each epoch at different learning rates in our training process with SET2 input, with more details in **Figure S7** (**Note 6 in SI**). From 2E-5 to 2E-4 learning rates, the time of one epoch is very close. This suggests that it takes much less time to reach optimal accuracy when increasing the learning rate. On the other hand, **Figure 6d** shows the comparison of conventional memristor (SET1, 1 LTD/LTP) and spintronic memristor (SET2, 8 LTD/LTP) at 1E-4 learning rate. Under this condition, an increased learning rate is not beneficial to the training process in SET1, as it causes divergence evidenced from many sharp peaks in accuracy. In **Figure S8** (**Note 7 in SI**), we demonstrate the increasing training stability as the number of data input increase from 1 to 16. Notably, in **Figure 6e** we compare the time to finish one epoch as a function of input data number. It appears that increasing the number of data input into the activation function does not have dramatic influence on the time to finish one epoch until 32 input datasets. Since increasing the number of input datasets from 2 (SET1) to 16 (SET2) does not increase the time required to complete one epoch, our results demonstrate that the spintronic memristor can achieve comparable accuracy in significantly less time than the conventional electrically driven memristors.

It should be noticed that the main difference between SET1 and SET2 is input dataset in the activation layer. It is crucial to understand whether the improved performance of our neural network is due solely to a larger input dataset or to the increased number of resistance states available. To verify this, we conducted a controlled experiment. We generated 7 more sets of data based on one set of LTD/LTP curve, introducing a 1% noise level to simulate the repeated measurement, and input them in the activation layer. **Figure S9** (**Note 8 in SI**) shows that such simulation still results in instability with a 1E-4 learning rate. This demonstrates that it is the multi-resistance states enabled by magnetic tuning that stabilize the neural network at higher learning rates and allow for faster training compared to conventional single-plasticity memristors. This can be explained by the fact that with more resistance states introduced into the activation layer, the activation functions exhibit smoother derivatives with less sharp transitions or discontinuities. The improved derivative can prevent abrupt changes in gradient flow that may cause instability in the simulation result.

As the extra resistance state comes from the TMR effect, we also studied the influence of TMR amplitude on the device performance. We generated 3 sets of 8 LTD/LTP curves with different TMR amplitudes of 10%, 20% and 100% and input them into our neural network. **Figure 6f** shows the recognition accuracy curves of three TMR conditions with 1E-4 learning rate. It can be observed that with an increased TMR amplitude, the training process becomes more stable at high learning rate. The accuracy at 100 Epoch is also larger for memristor with higher TMR effect. This implies that the performance of our magnetic tunable device can be further improved if we can introduce stronger TMR effect in the memristor.

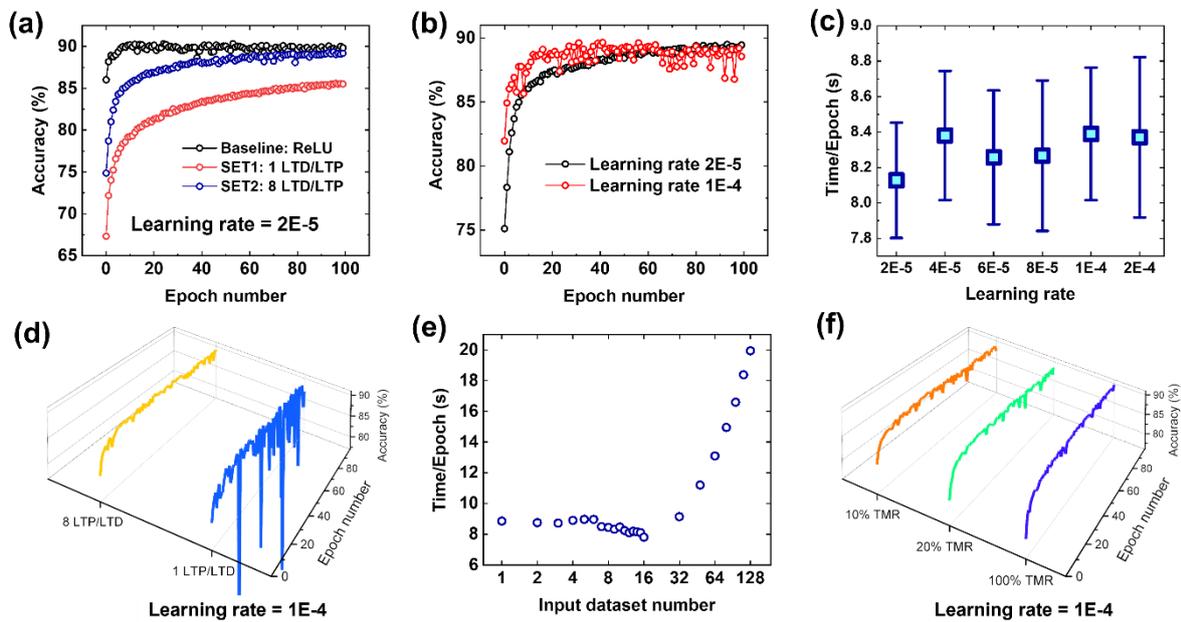

**Figure 6:** (a) Recognition accuracy as a function of epoch number for CNNs with different activation layer: ReLU (Baseline), 1 LTD/LTP curves at zero magnetic field (SET1) and 8 LTD/LTP (SET2) measured from our device. (b) Influence of learning rate on the recognition accuracy for CNN with EM activation layer of 8 LTD/LTP curves (SET2). With larger learning rate, the neural network can reach similar accuracy with fewer training epochs. (c) Average time for one epoch at different learning rates ranging from 2E-5 to 2E-4 with SET2 input. (d) Comparison of recognition accuracy

as a function of epoch number for CNNs with activation layer of 1 LTD/LTP (SET1) and 8 LTD/LTP (SET2) at high learning rate (1E-4). The CNN with 1 set of LTD/LTP curve input cannot be trained properly. (e) Average time for one epoch as a function of input dataset number. (f) Influence of TMR effect on the CNN performance. With larger TMR effect, the training becomes more stable and the accuracy at 100 epoch is higher.

*Discussions*

Memristors enable true in-memory computing by storing synaptic weights as analog resistance values directly within the device, drastically reducing data movement and latency, while at the same time enabling highly integrated system for large scale parallel computation, making them ideal candidates for scalable edge-AI platform. The interplay of TMR effect with artificial synapse behavior additionally brings several advantages in neuromorphic applications. The new degree of freedom to tune junction resistance by magnetic field makes it possible to achieve more resistance states in one single device. The increased resistance states improve the non-linearity and the gradient flow of the activation function, leading to better accuracy and faster training. In contrast, to achieve comparable performance, conventional electrically driven memristors typically require multiple devices[43] or carefully programmed pulses[44,45], increasing both the cost and complexity of the system. For large-scale integrated neuromorphic systems, multiple distinct LTD and LTP curves can be obtained by applying localized magnetic field (for example, a micro-coil surrounding the memristor[46], or a current line above the junction[47]). The tuning resistance by wireless magnetic field can avoid possible damage in memristor due to the Joule heating effect and electrical breakdown, which can significantly prolong the device lifetime. Also, the resistance modification from magnetic field can compensate the non-uniformity in individual devices, as well as the degradation of performance over time. To further reduce the energy consumption associated with magnetic field generation, alternative methods of electrical modulation of magnetization, such as spin-orbit torque via the spin Hall effect,[48] can also be utilized. More discussion can be found in **Note 10 in SI**.

**Conclusions**

In conclusion, we have demonstrated a novel strategy to improve the memristor performance by using magnetic field as a new degree of freedom to control the junction resistance. By integrating the TMR effect in the LSMO/PVDF/Co based organic spintronic memristor, we successfully achieve simultaneous control of memristor resistance using both electric and magnetic fields. In CNN inference and training, the magnetic-field-expanded resistance states achieve approximately 89% accuracy after 100 epochs—closely approaching the ReLU baseline (~90%) and outperforming the conventional electrically polarized memristor case (~84%). Moreover, they support faster and more stable training, enabling learning rates at least five times higher than those used with conventional electrically driven memristors. These results highlight the organic spintronic memristor as a more efficient and sustainable building block for integrated neuromorphic circuits, leveraging the flexibility of magnetization control. This paves the way for efficient, adaptable, and high-performance neuromorphic devices suitable for flexible applications in wearable, implantable, and other advanced electronic systems.

**Methods**

*Sample preparation*

The LSMO/PVDF/Co junction was fabricated by following procedure. A 50 nm $La_{0.67}Sr_{0.33}MnO_3$ (LSMO) film is grown on a (001)-oriented $SrTiO_3$ substrate via pulsed laser deposition, using a KrF (248 nm) laser. This deposition occurs at 750°C in a 300 mTorr oxygen atmosphere. The LSMO bottom electrode underwent UV lithography and ion milling to create junctions with sizes 10×10 μm². Polyvinylidene fluoride (PVDF) films were then deposited using the Langmuir-Blodgett method. A 0.01 wt% PVDF solution in dimethyl sulfoxide (DMSO) was first spread onto ultra-pure water. After compressing the PVDF molecules at the water surface to a surface pressure of 5 mN/m, a single PVDF layer was transferred onto the substrate by vertically lifting it at a speed of 3 mm/min, followed by drying for 30 minutes. The process was repeated three times to precisely deposit three layers of PVDF on LSMO bottom electrode. Following this, the PVDF film was annealed at 120ºC both in a glove box under an Ar atmosphere and in a molecular beam epitaxy (MBE) system under vacuum for 1 hour to eliminate the water and $CO_2$ contamination. High-resolution scanning transmission electron microscopy (HR-STEM) combined with electron energy loss spectroscopy (EELS) analysis clearly shows that the annealing process induces initial decomposition of the PVDF by the LSMO substrate.[21] After annealing, 10 nm Co and 10 nm Au layers were deposited on the PVDF using MBE to form the top electrodes. Due to residual oxygen present on the PVDF surface, the Co layer is partially oxidized, resulting in the formation of a ~4 nm thick CoO layer at the PVDF/Co interface.[21] After MBE deposition, UV lithography and ion milling were used again to pattern the top electrodes. To complete the structure, 10 nm of Ti and 150 nm of Au were deposited onto both the LSMO and Co/Au electrodes to serve as bonding pads.

*Magneto-transport measurements*

Magneto-transport measurements were performed using a closed-cycle helium cryostat, which provides a controlled low-temperature environment. The two-terminal *I-V* measurements were conducted using a Keithley 2450 as the voltage source and a Keithley 6487 picoammeter to measure the resulting current. To polarize the PVDF barrier, voltage pulses with duration of 1 s were applied to the junction with Keithley 2450. The pulses were applied with a ramp rate of 1 V/s and varied in amplitude. With help of Python programming, a Keithley 2400 unit was used to generate 10 ms pulses with various amplitude for neuron synaptic measurements.

*Neural network construction*

We train and test the neural network on the FashionMNIST dataset with 60k training images and 10k testing images.[49] For each image (28 × 28 × 3), we apply two blocks of convolutional layers to extract EM response features with the dimension of (4 × 4 × 16). Each block contains a Batch Normalization layer (its dimension equals to the channel size of the input), a 2-D convolutional layer (kernel size equals to 5), and an EM activation layer. The two blocks are sequentially connected to process the input image (as in the top row in **Figure 5a**). After obtaining the EM responses, we flatten them into a column vector and feed it to two blocks of MLP layers. Each block contains a Batch Normalization layer, an MLP layer, and an EM activation layer. The first MLP block reduces the dimension of EM responses from 256 × 1 to 120 × 1 followed by the second block reducing it to 84 × 1. Finally, we apply another set of Batch Normalization layer and MLP layer to map the features from 84 × 1 to 10 × 1 to calculate the classification score for each label. Note that 10 is the number of image labels in the FashionMNIST dataset. For each input image, there is a ground-truth class label L in {1,2,…,10}, which is used to train our network. We use the Cross Entropy function to calculate the loss for network training.

**Acknowledgements**

We acknowledge the help of Xavier Devaux for STEM-EELS characterizations. This work is supported by French National Research Agency (ANR) projects of FEOrgSpin (Grant No. ANR-18-CE24-0017), SOTspinLED (Grant No. ANR-22-CE24-0006-01) and PEPR SPIN SPINMAT (Grant No. ANR-22-EXSP-0007). The work at USTC was supported by National Natural Science Foundation of China (Grant No. 52422209). We also acknowledge ICEEL (international) PolarSPS project. Experiments were performed using equipment from the platform CC-Davm funded by

**Author contributions**

Y.L. coordinated the research project and designed the sample structure. S.S. and X.L. prepared STO/LSMO substrates. Y.H. and T.C. fabricated devices with UV lithography. T.C. and Y.H. characterized magneto-transport properties. T.C. characterized the junction synaptic properties. Y.N., T.C. and J.P. contributed to neural network simulation. T.C., Y.L. and Y.N. prepared the manuscript. All authors analyzed the data, discussed the results and commented on the manuscript.

**Supporting information**

Morphology characterization of PVDF on LSMO; STEM-EELS characterization of LSMO/PVDF/Co junction after cyclic polarization; optimization of polarization voltage amplitude for clear observation of LTP and LTD behaviors; retention characterization of LSMO/PVDF/Co memristor; optimization of pre- and post-synaptic spikes for STDP measurements; analysis of epoch duration at different learning rates; effect of input LTP/LTD curve number on recognition stability; influence of the number of input dataset on the recognition accuracy; loss function for different activation function settings and different learning rates; discussion on energy consumption for magnetic field control.

**TOC figure**

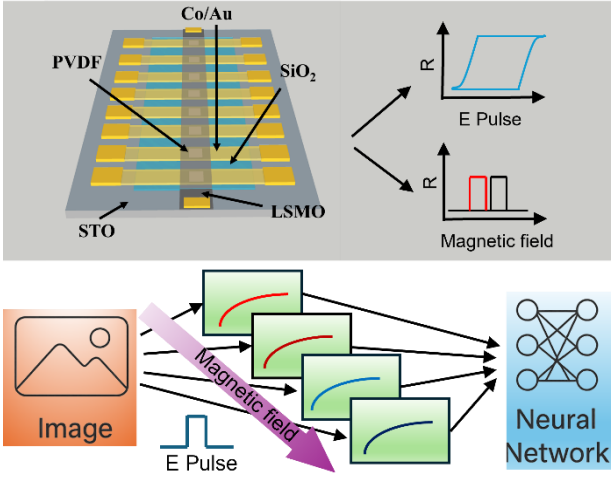

# Supporting Information

**Magnetic-field controlled organic spintronic memristor for neural network computation**

Tongxin Chen[1], Yinyu Nie[2], Yafei Hao[1,5], Shengchun Shen[3], Jiajun Pan[4], Xiaoguang Li[3], Yuan Lu[1*]

[1]*Université de Lorraine, CNRS, Institut Jean Lamour, F-54000 Nancy, France*
[2]*Technical University of Munich, Arcisstraße 21, 80333 München, Germany*
[3]*Hefei National Research Center for Physical Sciences at the Microscale, Department of Physics, University of Science and Technology of China, Hefei 230026, China*
[4]*University of Lorraine, Laboratoire Lorrain de Recherche en Informatique et ses Applications (LORIA), UMR 7503, 54506 Vandoeuvre-les-Nancy, France*
[5]*Department of Physics, Zhejiang Normal University, Jinhua 321004, China*

Corresponding authors*: *yuan.lu@univ-lorraine.fr*

# Table of content

**Note 1: Morphology characterization of PVDF on LSMO**

To assess the uniformity of the ultrathin PVDF film prepared by the Langmuir–Blodgett (LB) method, we performed atomic force microscopy (AFM) measurements on a five-layer PVDF film deposited on an LSMO substrate. As shown in **Fig. S1**, the surface exhibits a relatively uniform morphology over a 30 × 30 μm² area with a root-mean-square (RMS) surface roughness of about 1.3 nm. However, some flake-like PVDF domains can be observed, which is a typical feature of LB films due to the partial surface coverage in each LB deposition cycle. In future work, we aim to further improve uniformity by optimizing the surface tension and transfer rate during the LB process.

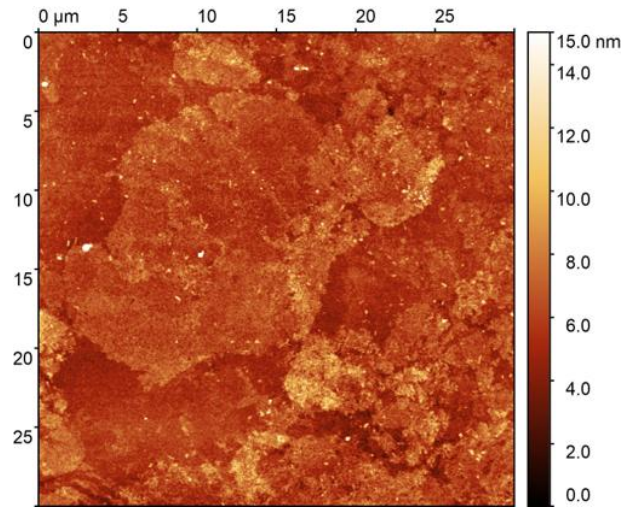

**Figure S1**: AFM topography measurement of the PVDF surface over 30×30 μm² area on STO/LSMO (50 nm) substrate. Adapted from Ref. [1].

**Note 2: STEM-EELS characterization of LSMO/PVDF/Co junction after cyclic polarization**

To better understand the polarization mechanism, we employed high-resolution scanning transmission electron microscopy (HR-STEM) to characterize the interfacial structures, combined with spatially-resolved electron energy loss spectroscopy (EELS) to analyze the chemical distribution within the junction. **Fig. S2a** presents a STEM image of the junction after 100 cycles of polarization (alternating positive and negative polarization, ending with negative polarization). The different layers are clearly distinguishable. The thickness of the PVDF layer ranges from 2 to 5 nm. As reported in our previous study,[1] the F elements are completely decomposed from the PVDF layer after annealing at 120 °C by the LSMO bottom electrode. Fluorine atoms are expelled from the PVDF layer, moving toward the LSMO side under positive polarization or toward the Co side under negative polarization.

**Figs. S2c–h** show elemental maps obtained from processed EELS spectra. Due to oxygen presence at the PVDF/Co interface during sample preparation, the Co in contact with PVDF undergoes partial oxidation, forming a ~4 nm thick CoO layer. Interestingly, fluorine is not only distributed within this CoO layer but is also present at the LSMO/PVDF interface (**Fig. S2g**). **Fig. S2b** shows the elemental profiles across the PVDF junction, offering a clearer view of elemental redistribution after cyclic polarization. The fluorine profile exhibits two distinct peaks: the majority of F atoms (>75%) are located within the CoO layer, overlapping significantly with the oxygen distribution in Co, while a smaller fraction (~25%) remains at the LSMO/PVDF interface.

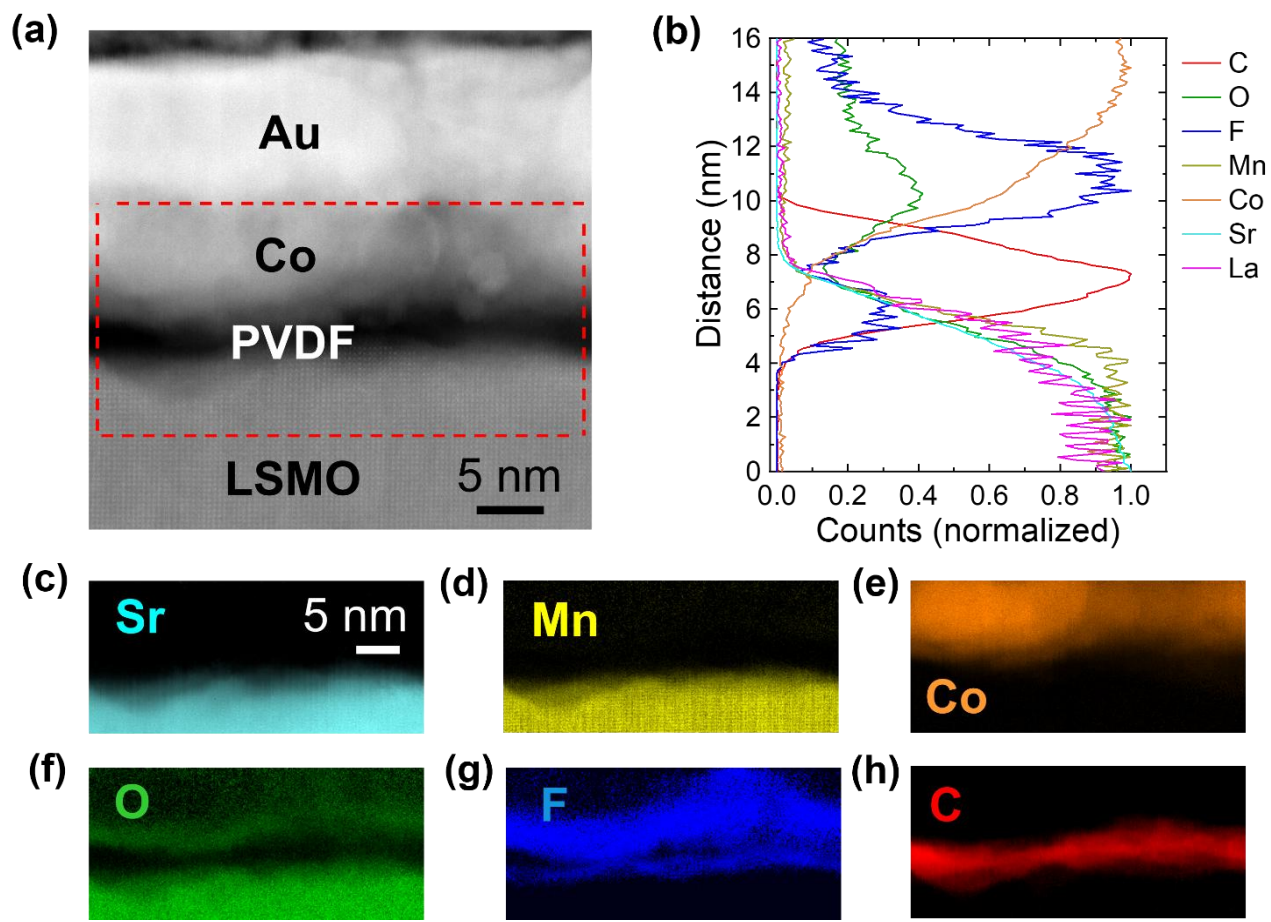

**Figure S2:** Interfacial structure and chemical characterization of the LSMO/PVDF/Co/Au junction after cyclic polarizations. (a) STEM image of interface structure of different layers. Red dashed square shows the area where EELS maps in (c-h) were extracted. (b) Normalized elemental profiles from LSMO to Co drawn from EELS elemental maps. (c-h) EELS elemental maps for (c) Strontium, (d) Manganese, (e) Cobalt, (f) Oxygen, (g) Fluorine, and (h) Carbon. Adapted from Ref. [1].

**Note 3: Optimization of polarization voltage amplitude for clear observation of LTP and LTD behaviors**

To observe both long-term potentiation (LTP) and long-term depression (LTD) behavior in our junction, careful selection of the polarization voltage amplitude is crucial. Applying excessively high negative pulses drives the junction rapidly into a low-resistance state (LRS), thereby obscuring the LTP behavior (**Fig. S3a**). Near the LRS region, where F atoms have already entered the CoO layer, further fluorine incorporation causes small additional changes in resistance because F-doped CoO becomes highly conductive.[1] When we reduce the amplitude of the negative pulses to –1 V (**Fig. S3b**), the CoO remains less conductive, and the junction resistance gradually decreases with the number of pulses. Further lowering the negative pulse amplitude to -0.8 V allows us to observe clear LTP behavior, as shown in **Fig. S3c**.

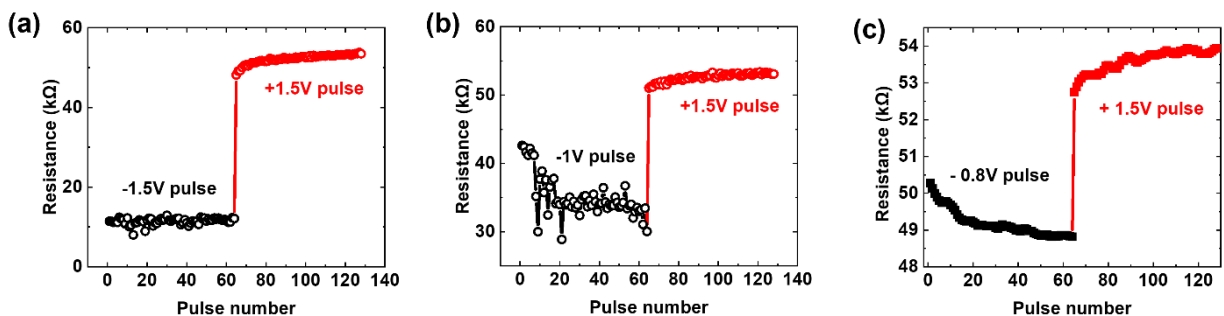

**Figure S3:** LTP/LTD measurement with different pulse amplitudes of (a) -1.5 V/+1.5 V, (b) -1.0 V/+1.5 V, and (c) -0.8 V/+1.5 V, respectively. The duration of all pulses is 10 ms.

To further enhance the resistance variation, the amplitude of the positive pulses can be increased, driving F out of the CoO layer toward the PVDF/LSMO interface and restoring the high resistance state (HRS).[1] As shown in **Fig. S4**, increasing the positive polarization voltage from +1.5 V to +2 V results in a higher resistance state and a larger resistance variation during the LTD process. The enhanced positive polarization subsequently enables the application of a larger negative polarization voltage of -1.3 V, leading to a more stable and greater resistance variation during the LTP process.

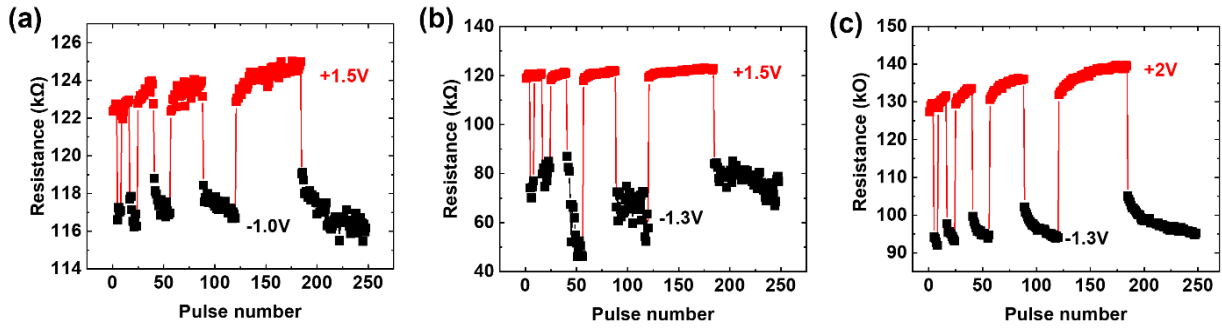

**Figure S4**: Repeated LTD/LTP measurement with different pulse amplitudes of (a) +1.5 V/-1.0 V, (b) +1.5 V/-1.3 V and (c) +2 V/-1.3 V, respectively. The duration of all pulses is 10 ms.

# Note 4: Retention characterization of LSMO/PVDF/Co memristor

To assess the endurance of resistive switching, we conducted 100 cycles of alternate +1.8 V and -2.0 V polarizations on the junction, as demonstrated in **Fig. S5a**. Both resistance states are well preserved after 100 polarization cycles. To characterize the retention time of LSMO/PVDF/Co memristor, we have polarized the junction into two resistance states and kept them for 100 minutes. As shown in **Fig. S5c**, both polarization states have a good stability, suggesting that the F element can stably reside at each interface once the junction is polarized.

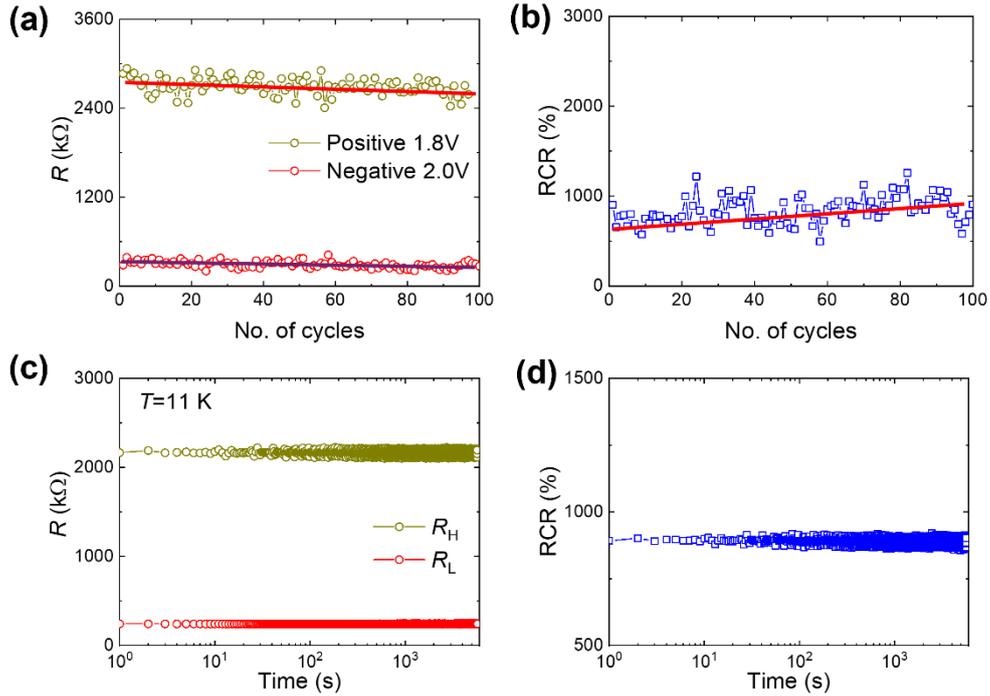

**Figure S5:** (a) 100 cycles of resistive switching at $T$=11 K with $V_P$=+1.8 V (dark yellow) and $V_P$=−2.0 V (red). (c) The resistance retention time is measured for 100 minutes after one polarizing pulse with $V_P$=+1.9 V (dark yellow) and $V_P$=−1.9 V (red), respectively. (b,d) Resistance change ratio (RCR) calculated from the resistance change presented in (a,c), respectively. Adapted from Ref. [1].

**Note 5: Optimization of pre- and post-synaptic spikes for STDP measurements**

In our STDP measurements, the selection of presynaptic and postsynaptic pulse amplitudes and widths was based on a combination of empirical optimization and the need to achieve stable and biologically relevant long-term potentiation (LTP) and depression (LTD) behaviors.

As shown in **Fig. S3**, larger negative pulses (e.g., −1.5 V combined with +1.5 V) tend to drive the device quickly into a low-resistance state, suppressing gradual LTP and reducing dynamic range. In contrast, reducing the negative amplitude to −1.0 V or −0.8 V allows for smoother, incremental potentiation and depression behaviors. Further optimization is discussed in **Fig. S4**, where we observed that increasing the positive amplitude to +2.0 V could help recover the high-resistance state and expand the usable conductance range. Under this condition, a slightly higher negative amplitude (e.g., −1.3 V) can also be used while maintaining device stability. From these observations, we conclude that the device's effective threshold for conductance change lies around +2.0 V (positive) and −1.3 V (negative).

When the pre- and post-synaptic spikes overlap within a certain time window (Δt), their combined waveform can temporarily exceed the device's threshold voltage, triggering a change in conductance. To ensure that this overlap leads to controlled and gradual weight updates (rather than abrupt switching or saturation), we carefully tuned the pulse parameters. Specifically, as shown in **Fig. S6a**, we selected the amplitude combinations of (+1.0 V/−0.67 V) and (+0.67 V/-1.0 V) for pre- and post-synaptic spikes, respectively, based on prior experimental experience of conductance change threshold (+2.0 V/-1.3 V). To faithfully simulate STDP, we calculated the expected waveform overlaps using a Python-based model and then applied them to the device using a Keithley 2400 source meter. In **Fig. S6b**, we show the combined pre- and post-synaptic waveforms for Δt = ±10 ms. Within a short time window, the combined pulse amplitude can reach the conductance change threshold (+2.0 V/-1.3 V) to effectively modify the memristor conductance.

Due to the limitations of the Keithley 2400, the minimum pulse duration we could reliably apply was 5 ms. Within this constraint, the chosen amplitudes were optimized to avoid saturation while still enabling clear LTP and LTD responses. We acknowledge that other combinations (such as higher amplitudes or shorter pulses) could potentially improve the time resolution of STDP emulation. However, in our current setup, increasing the amplitude or duration tends to accelerate device saturation and reduce the number of effective conductance states. Shorter pulse durations (e.g., sub-millisecond) would require higher amplitudes to exceed the threshold within the reduced time, which we plan to explore in future work with higher-speed pulse generators.

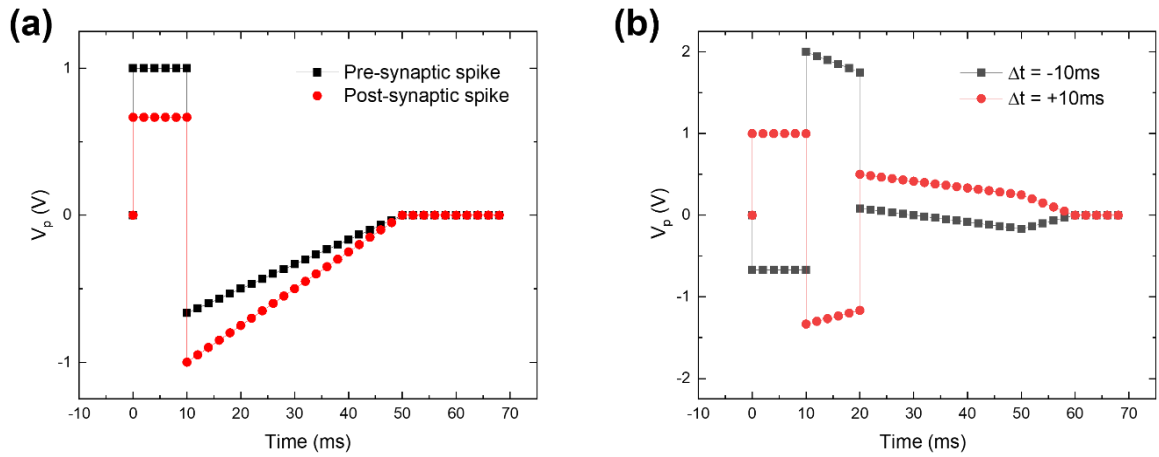

**Figure S6:** (a) pre- and post-synaptic spikes with a total length of 50 ms. The amplitudes of the rectangular voltage pulse and opposite slope are +1.0 and −0.67 V for the spike from pre-synaptic neuron, and +0.67 and −1.0 V for the spike from post-synaptic neuron, respectively. (b) Waveforms resulting from superposition of pre- and post-synaptic spikes when Δt=-10ms and  or Δt=+10ms. The post-neuron spike is inverted for generating the combined waveforms.

**Note 6: Analysis of epoch duration at different learning rates**

To investigate the training time in our simulation, we recorded the duration of each epoch during training with different learning rates. **Fig. S7** shows the time of one epoch as a function of epoch number at 2E-5 and 1E-4 learning rates. The average times of each epoch under these two conditions are presented in **Figure 6c in main text**.

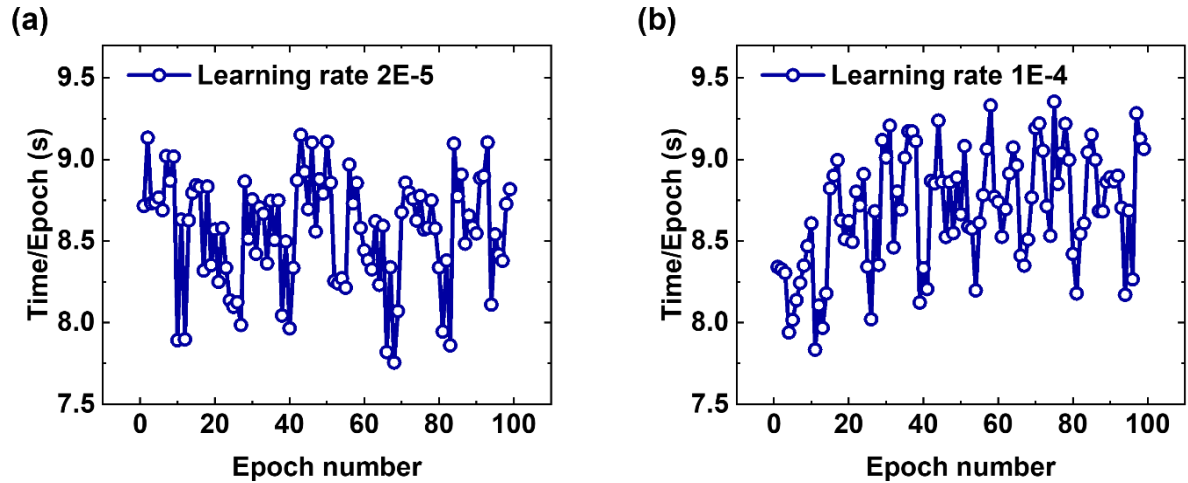

**Figure S7:** Time of one epoch as a function of epoch number at (a) 2E-5 and (b) 1E-4 learning rates.

**Note 7: Effect of input LTP/LTD curve number on recognition stability**

We conducted a controlled experiment by inputting 1 to 16 LTP and LTD curves into the activation function, using data obtained from our device. As clearly shown in **Fig. S8**, at a learning rate of 1E-4, the recognition gradually became more stable during training as the number of input LTP/LTD curves increased.

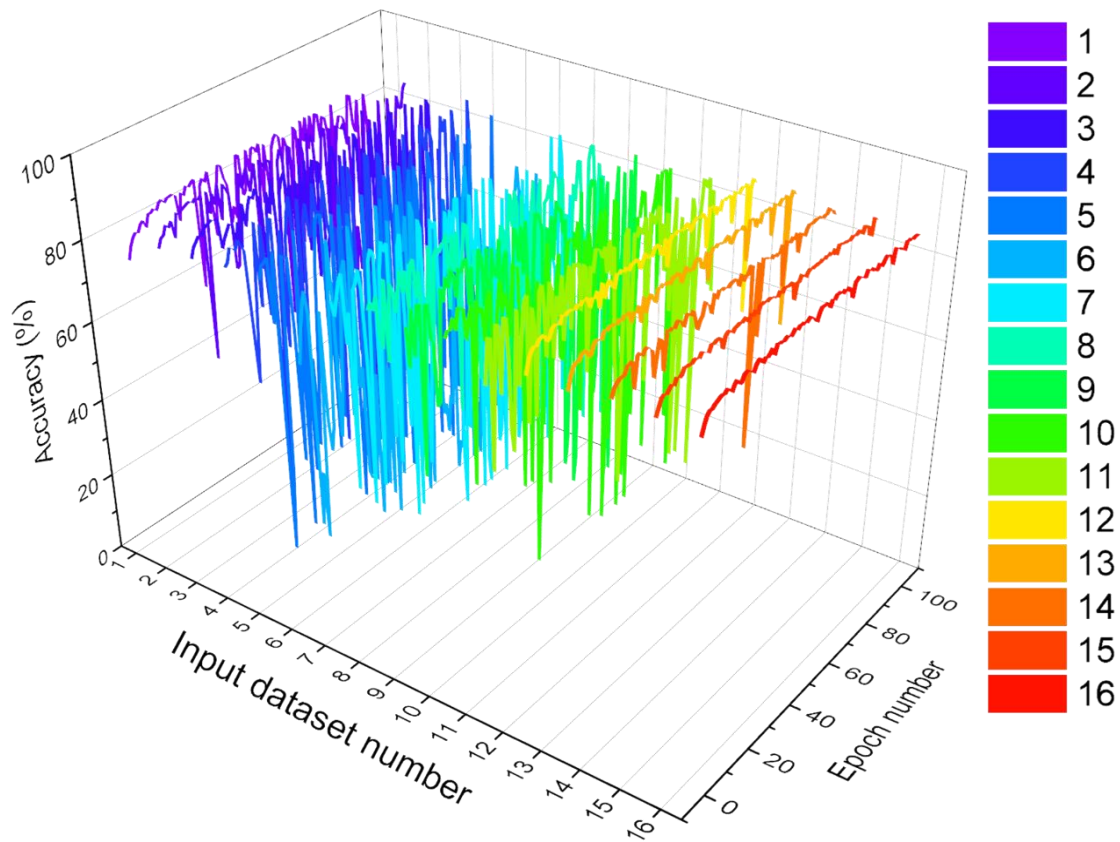

**Figure S8:** The evolution of recognition accuracy during training as a function of input LTP/LTD curve number.

**Note 8: Influence of the number of input dataset on the recognition accuracy**

To evaluate the influence of the number of input dataset on the recognition accuracy, we conducted controlled experiments. Starting from one set of LTP/LTD curves, we generated seven additional datasets by introducing 1% noise to simulate repeated measurements, resulting in a total of 16 curves as inputs to the activation layer. With a low learning rate (2E-5) (**Fig. S9a**), the training results are similar to those obtained using only one set of LTP/LTD curves, and the accuracy is lower than that achieved when using 8 sets of LTP/LTD curves measured by varying magnetic field. At a higher learning rate (1E-4) (**Fig. S9b**), the training performance becomes instable when using the noised curves, in contrast to the improved stability observed with 8 sets of LTP/LTD curves. These results exclude the possibility that the improved robustness stems simply from an increased amount of input dataset.

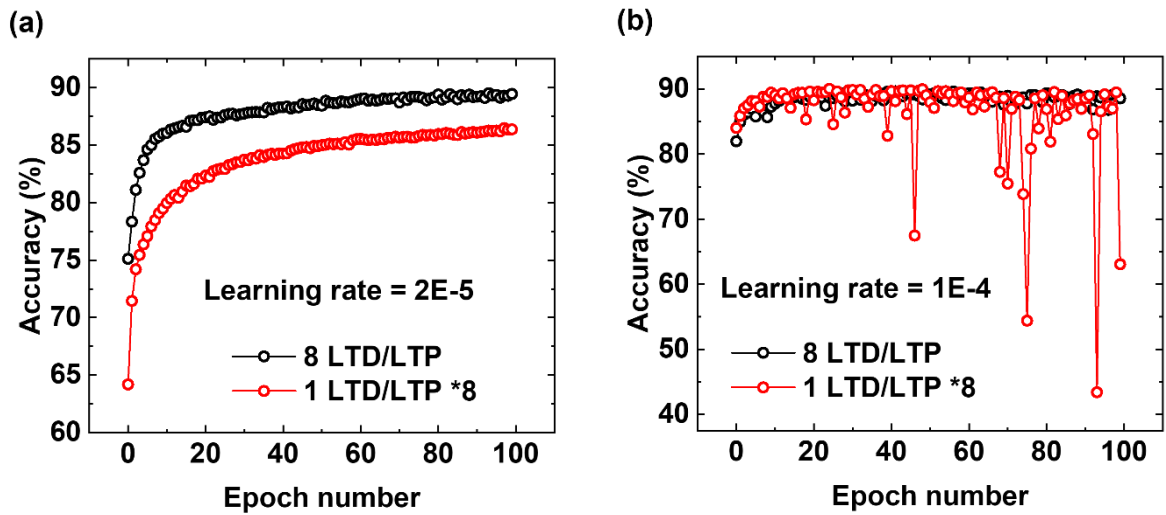

**Figure S9:** Influence of input dataset number on the recognition accuracy at (a) 2E-5 and (b) 1E-4 learning rates.

**Note 9: Loss function for different activation function settings and different learning rates**

In **Fig. 6d in main text**, we compare the recognition accuracy as a function of epoch number for CNNs using two different activation settings: 1 LTD/LTP pair (SET1) and 8 LTD/LTP pairs (SET2), both trained with a high learning rate of $1\times10^{-4}$. We show the evolution of the loss function during training for both configurations in **Fig. S10a,b**. Both loss curves exhibit a clear downward trend and converge stably without sustained divergence. The 16-data configuration (SET2, **Fig. S10a**) does not show any evidence of persistent "local-minimum jumps." Occasional, brief upward steps do occur, but they are rare, transient, and typically reconverge within a few iterations—behaviors consistent with minibatch stochasticity rather than true basin-to-basin transitions in the loss landscape. Because the learning rate is fixed in **Fig. 6d**, the improved stability observed with SET2 can be attributed to the activation function rather than to the learning rate itself. Specifically, the 16-data activation appears to induce a smoother and better-conditioned nonlinearity, which likely reduces batch-to-batch prediction variability and contributes to the steadier accuracy curve.

We further compare the convergence behavior under two different learning rates ($1\times10^{-4}$ and $2\times10^{-5}$) using the same network architecture and 16-data activation. As shown in **Fig. S10c,d**, The higher learning rate achieves faster convergence, with the loss function dropping more quickly and stabilizing at a lower plateau (**Fig. S10c**). In contrast, the lower learning rate results in a smoother per-step progression, but convergence is significantly slower and terminates at a higher final loss (**Fig. S10d**). In both cases, we observe no signs of persistent local minima trapping. The faster convergence and lower terminal loss of the higher learning rate suggest better training efficiency. For further improvement, a mild learning rate decay in later epochs can help suppress small residual oscillations without compromising convergence speed.

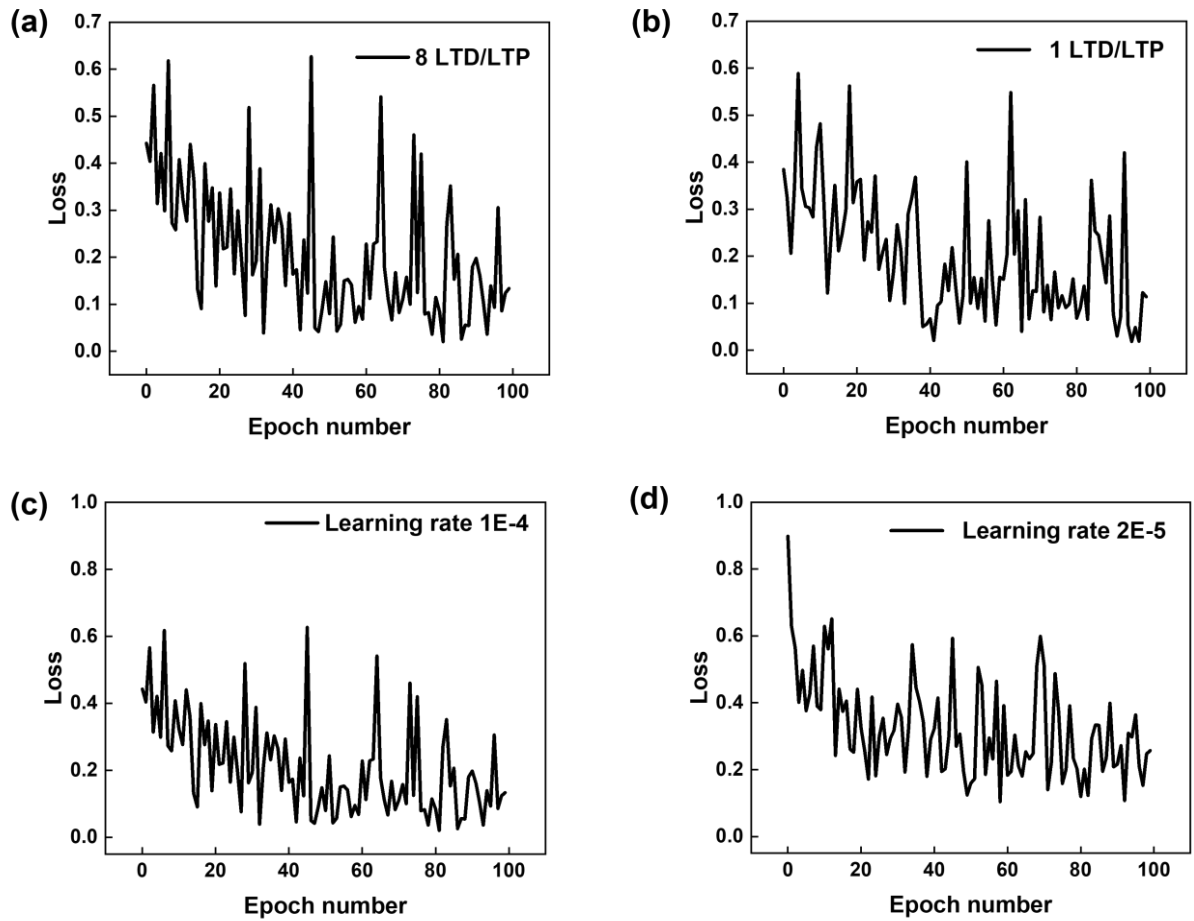

**Figure S10**: (a,b) The evolution of the loss function during training for two different activation settings: (a) 8 LTD/LTP pair (SET2) and (b) 1 LTD/LTP pairs (SET1), both trained with a high learning rate of 1E-4. (c,d) The evolution of the loss function during training for two different learning rates: (c) 1E-4 and (d) 2E-5 using the same network architecture of 16-data activation (SET2).

**Note 10: Discussion on energy consumption for magnetic field control**

The energy consumption associated with magnetic field control must be carefully considered for practical applications, particularly when comparing it with purely electrical methods. In a simple approach, the magnetic field can be locally generated using an on-chip electrical line placed in close proximity to the device (approximately ~20 nm). For example, passing a 1 mA current through a gold line with dimensions of 1 mm length, 1 μm width, and 100 nm thickness (estimated resistance ~100 Ω) can produce a magnetic field of approximately 100 G, which is sufficient to modulate the junction resistance. If this magnetic field is only applied briefly during the weight update process (typically ~1 μs, though the magnetization switching itself occurs in less than 1 ns), the energy consumed per update is approximately: $E=I^2Rt=(1mA)^2\times100\Omega\times1\mu s=0.1nJ$.

While this does introduce an additional energy overhead compared to purely electrical control, we find that fewer training epochs are needed to reach a target accuracy when using magnetic-field-assisted states. Thus, the cumulative energy consumption during the training process can still be lower overall, due to the improved convergence behavior.

Additionally, it is possible to supply the magnetic field via passive means, such as by integrating a permanent magnetic layer (e.g., NiFe) that produces a built-in magnetic field gradient. By arranging the devices spatially around such a magnetic source, each device can experience a different local field without any ongoing energy cost for magnetic field generation[2]. In this scenario, the energy associated with magnetic field control becomes negligible, enabling highly energy-efficient operation.

Looking forward, we are also exploring all-electrical magnetization switching mechanisms, such as using spin-orbit torque (SOT) via the spin-Hall effect in ferromagnet/heavy-metal (FM/HM) bilayer structures[3]. In such mechanism, a charge current through the heavy metal layer generates a transverse spin current that exerts a torque on the adjacent ferromagnetic layer, enabling deterministic and reversible switching of magnetization. This method is not only compatible with CMOS fabrication but also offers ultra-fast (sub-nanosecond) switching with extremely low energy. For instance, using a 10 mA, 1 ns pulse through a 100 Ω electrode results in an energy cost of only: $E=I^2Rt=(10mA)^2\times100\Omega\times1ns=0.01nJ$.

Furthermore, the magnetization state achieved via SOT is non-volatile, allowing the device to retain its state without standby power. This offers significant advantages in terms of scalability, energy efficiency, and compatibility for future high-density neuromorphic or memory-in-compute systems.

In conclusion, while magnetic field control introduces some initial energy costs, careful device design—through local on-chip magnetic field generation, passive magnetic sources, or spintronic mechanisms like SOT—can enable highly energy-efficient and scalable systems. Moreover, spintronic devices with all-electrical control are fully compatible with dense integration and hold great promise for next-generation, high-compatibility neuromorphic computing applications.